\documentclass[9pt, technote]{IEEEtran}
\usepackage{amsmath}   % From the American Mathematical Society
\usepackage{amsfonts}
\usepackage{amssymb}
\usepackage{latexsym}
\newtheorem{theorem}{Theorem}[section]
\newtheorem{remark}[theorem]{Remark}

\newtheorem{lemma}[theorem]{Lemma}
\newtheorem{corollary}[theorem]{Corollary}
\newcommand{\F}{{\mathbf F_2}}

\newcommand{\Symp}{{\mathrm{Sp}}}

\newcommand{\LU}{{\mathrm{LU}}}

\newcommand{\diag}{{\mathrm{diag}}}

\newcommand{\abs}[1]{|#1|}
\begin{document}
\title{On the dimensions of certain LDPC codes based on $q$-regular bipartite graphs}
\author{Peter Sin and Qing Xiang%
\thanks{P. Sin is with the Mathematics Department at the University of Florida.}%
\thanks{Q. Xiang is with the Mathematics Department at the University of Delaware. His work was supported by NSF grant DMS0400411.}}
\markboth{}{Sin \MakeLowercase{\textit{et al.}}: LU codes}
\maketitle
\begin{abstract}
An explicit construction of a family of binary 
LDPC codes called $\LU(3,q)$, where $q$ is a power
of a prime, was recently given.
A conjecture was made for the dimensions of these codes
when $q$ is odd.
The conjecture is proved in this note.
The proof involves the geometry  
of a $4$-dimensional symplectic vector space and 
the action of the symplectic group and its subgroups.
\end{abstract}
\IEEEpeerreviewmaketitle

%}}}
%{{{ body

\section{Introduction}

%\PARstart{L}{et} $V$ be a $4$-dimensional
Let $V$ be a $4$-dimensional
vector space over the field $\mathbf F_q$ of $q$ elements. 
We assume that $V$ has a nonsingular alternating 
bilinear form $(v,v')$ and
denote by $\Symp(V)$ the group of linear automorphisms of $V$
which preserve this form. We
 choose a symplectic basis  $e_0$, $e_1$, $e_2$, $e_3$ 
 of $V$, with $(e_i,e_{3-i})=1$, for $i=0$, $1$.

Let $P=\mathbf P(V)$
be the set of points of the projective space of $V$.
A subspace of $V$ is said to be \emph{totally isotropic}
if $(v,v')=0$ whenever $v$ and $v'$ are both in the
subspace. 
Let $L$ denote the set of totally isotropic 
$2$-dimensional subspaces of $V$, considered as lines in $P$.
The pair $(P,L)$, together with the natural relation of incidence
between points and lines, is called the {\it symplectic generalized quadrangle}. Except for in the appendix, the term ``line''
will always mean an element of $L$.
It is easy to verify that $(P,L)$ satisfies
the following {\it quadrangle property}:
Given any line and any point not on the line, 
there is a unique line which passes through
the given point and meets the given line.

Now fix a point $p_0\in P$ and a line ${\ell_0}\in L$ through $p_0$.
We can assume that we chose our basis so that 
$p_0=\langle e_0\rangle$ and $\ell_0=\langle e_0, e_1\rangle$.
For $p\in P$, denote by $p^\perp$ the set of points
on lines through $p$; $p'\in p^\perp$ if and only if
the subspace of $V$ spanned by $p$ and $p'$ is isotropic.
Consider the set $P_1=P\setminus p_0^\perp$ of points
not collinear with $p_0$, and the set $L_1$
of lines which do not meet ${\ell_0}$.
Then we can also consider the incidence systems $(P_1,L_1)$,
$(P,L_1)$ and $(P_1,L)$. 
Let $M(P,L)$ and $M(P_1,L_1)$
be the binary incidence matrices of the respective incidence systems,
with rows indexed by points and columns by lines.
The rows and columns of $M(P,L)$ have weight $q+1$ and,
as a consequence of the quadrangle property, those of
$M(P_1,L_1)$ have weight $q$. 

If $q$ is odd we know by Theorem 9.4 of \cite{Bagchi:lu}
that the $2$-rank of $M(P,L)$ is 
$(q^3+2q^2+q+2)/2$.
Here we prove the following theorem.
\begin{theorem} Assume $q$ is a power of an odd prime.
The $2$-rank of $M(P_1,L_1)$ equals $(q^3+2q^2-3q+2)/2$.
\label{main}
\end{theorem}

In \cite{Kim:lu}, a family of codes designated $\LU(3,q)$ was defined
in the following way. Let $P^*$ and $L^*$ 
be sets in bijection with ${\mathbf F_q}^3$, where $q$ is any
prime power. An element
$(a,b,c)\in P^*$ is incident with an element $[x,y,z]\in L^*$ if and only if
\begin{equation}\label{kimeqn}
y=ax+b \qquad\text{and}\qquad z=ay+c.
\end{equation}
The binary incidence matrix with rows indexed by $L^*$
and columns indexed by $P^*$ is denoted by $H(3,q)$ 
and the two binary codes having $H(3,q)$ and its transpose
as parity check matrices are called $\LU(3,q)$ codes.
The name comes from \cite{LU:lu}, where the
bipartite graph with parts $P^*$ and $L^*$ and
adjacency defined by the equations (\ref{kimeqn}) had
been studied previously.

It is not difficult to show that the incidence systems
$(P_1, L_1)$ and $(P^*, L^*)$  are equivalent. 
A detailed proof is given in the appendix.
Thus, $M(P_1,L_1)$ is a parity check matrix of the $\LU(3,q)$
code given by the transpose of $H(3,q)$ and 
Theorem~\ref{main} has the following immediate corollary.

\begin{corollary} If $q$ is a power of an odd prime, 
the dimension of $\LU(3,q)$ is 
 $(q^3-2q^2+3q-2)/2$.
\label{maincor}
\end{corollary}

The corollary was conjectured in \cite{Kim:lu}. There it was established that this number is a lower bound when $q$ is an odd prime.

\section{Relative dimensions and a lower bound for $\LU(3,q)$}
In this section $q$ is an arbitrary prime power.

Let $\F[P]$ be the vector space of
all $\F$-valued functions on $P$. We can think of such a function
as a vector in which the positions are indexed by the points of $P$,
and the entries are the values of the function at the points.
For $p\in P$, the characteristic function $\chi_p$ is the 
vector with $1$ in the position with index $p$
and zero in the other positions. 
The set of all characteristic functions
of points forms a basis of $\F[P]$.
%added some here
Let $\ell\in L$. Its characteristic function $\chi_\ell\in \F[P]$ 
is the function which takes the value $1$ at the $q+1$ 
points of $\ell$ and zero at all other points. 
The subspace of $\F[P]$ spanned
by all the $\chi_\ell$ is  the $\F$-code of $(P,L)$, denoted by $C(P,L)$.
One can think of $C(P,L)$ as the column space of $M(P,L)$. 
For brevity, we will sometimes blur the distinction between
lines and their characteristic functions and speak, for instance,
of the subspace of $\F[P]$ spanned by a set of lines. 
Let $C(P,L_1)$  be the subspace of $\F[P]$ spanned by 
lines in $L_1$. Let $C(P_1,L_1)$ denote the code of $(P_1, L_1)$,
viewed as a subspace of $\F[P_1]$, and let
$C(P_1,L)$  be the larger subspace of $\F[P_1]$ spanned by
the restrictions to $P_1$ of the characteristic functions of
all lines of $L$.

Consider the natural projection map
\begin{equation}
\pi_{P_1}: \F[P]\rightarrow \F[P_1]
\end{equation}
given by restriction of functions. Its kernel
will be denoted by $\ker\pi_{P_1}$.

Let $Z\subset C(P,L_1)$ be a set of characteristic
functions of lines in $L_1$ which 
maps bijectively under $\pi_{P_1}$ to a basis of $C(P_1, L_1)$.
Let $X$ be the set of characteristic functions
of the $q+1$ lines of $L$ through $p_0$ 
and let $X_0=X\setminus\{\chi_{\ell_0}\}$.
Finally, choose any $q$ lines of $L$ which  meet $\ell_0$ 
in the $q$ distinct points other than $p_0$ and
let $Y$ be the set of their characteristic functions.
It is clear that the sets $X$, $Y$ and $Z$ are disjoint
and that $X$ is contained in  $\ker\pi_{P_1}$.

\begin{lemma}\label{linindep} 
$Z\cup X_0\cup Y$ is linearly independent over $\F$.
\end{lemma} 
\begin{proof}
Each element of $Y$
contains in its support a point of $\ell_0$ which
is not in the support of any other element of  
$Z\cup X_0\cup Y$. So it is enough to show that
$X_0\cup Z$ is linearly independent. This is true because
$X_0$ is a linearly independent subset of  $\ker\pi_{P_1}$
and $Z$ maps bijectively under $\pi_{P_1}$
to a linearly independent set.
\end{proof}

We note that 
$\abs{Z}=\dim_\F C(P_1,L_1)$ and $\abs{X_0\cup Y}=2q$. 

\begin{corollary}\label{LUlowerbound} Let $q$ be an arbitrary
prime power. Then
\begin{equation}\label{ineq}
\dim_\F \LU(3,q) \geq q^3-\dim_\F C(P,L)+2q.
\end{equation}
\end{corollary}
\begin{proof}
From the definition of $\LU(3,q)$ and the equivalence
of $(P^*,L^*)$ with $(P_1, L_1)$,
we have 
\begin{equation}
\dim_\F \LU(3,q)=q^3-\dim_\F C(P_1,L_1).
\end{equation}
The corollary
now follows from Lemma~\ref{linindep}.
\end{proof}
\section{Proof of Theorem~\ref{main}}
In this section we assume that $q$ is odd.
In view of Corollary~\ref{LUlowerbound} and the known
$2$-rank of $M(P,L)$ 
the proof of  Theorem~\ref{main} will be completed if we can
show that $Z\cup X_0\cup Y$  spans $C(P,L)$ as a vector
space over $\F$. 

\begin{lemma}\label{allone} Let $\ell\in L$.
Then the sum of the 
characteristic functions of all lines which meet $\ell$ (excluding
$\ell$ itself) is the constant function 1.
\end{lemma}
\begin{proof}  
The function given by the sum takes the value
$q\equiv 1$ $({\rm mod\ } 2)$ at any point of $\ell$ and value $1$ at any point off $\ell$, by the quadrangle property.
\end{proof}

\begin{lemma}\label{Phi} Let $\ell\in L$ be a line, other 
than $\ell_0$, which meets $\ell_0$ at a point $p$.
Let $\Phi_\ell$ be the sum of all the characteristic
functions of lines in $L_1$ which meet $\ell$.
Then
\begin{equation}
\Phi_\ell(p')=\begin{cases} 0, \quad\text{if $p'=p$};\\
                        q, \quad\text{if $p'\in \ell\setminus\{p\}$};\\
                        0, \quad\text{if $p'\in p^\perp\setminus\ell$};\\
                        1, \quad\text{if $p'\in P\setminus p^\perp$}.
\end{cases}
\end{equation}
\end{lemma}
\begin{proof}This is an immediate consequence of
the quadrangle property.
\end{proof}
\begin{corollary}\label{linediffs} Let $p\in \ell_0$ and let $\ell$, $\ell'$
be two lines through $p$, neither equal to $\ell_0$.
Then $\chi_\ell-\chi_{\ell'}\in C(P, L_1)$.
\end{corollary}
\begin{proof} Since $q=1$ in $\F$, one easily check
using Lemma~\ref{Phi} that
\begin{equation}
\chi_\ell-\chi_{\ell'}=\Phi_\ell-\Phi_{\ell'}\in C(P, L_1).
\end{equation}
\end{proof}

We now come to our main technical lemma.

\begin{lemma}\label{kerP} $\ker\pi_{P_1}\cap C(P,L)$ has dimension $q+1$, with basis the set $X$ of characteristic functions
of the $q+1$ lines through $p_0$.
\end{lemma}
\begin{proof}
Let $G_{p_0}$ be the stabilizer in $\Symp(V)$ of $p_0$.

From the definition,
\begin{equation}
\ker\pi_{P_1}=\F[p_0^\perp]=\F[\{p_0\}]\oplus\F[{p_0}^\perp\setminus\{p_0\}]
\end{equation}
as an $\F G_{p_0}$-module. Clearly $\F[\{p_0\}]$ is a one-dimensional
trivial $\F G_{p_0}$-module. To find the structure of
$\F[{p_0}^\perp\setminus\{p_0\}]$, we consider the 
following subgroups of $G_{p_0}$, which we will describe
as matrix groups with respect to our chosen basis.

Let
\begin{equation}\label{Qmatrix}
Q=\lbrace\left(\begin{matrix}1&a&b&c\\
                       0&1&0&b\\
                       0&0&1&-a\\
                       0&0&0&1 \end{matrix}\right) \mid a,b,c\in\mathbf F_q\rbrace
\end{equation}
and
\begin{equation}
C=\lbrace\left(\begin{matrix}1&0&0&c\\
                       0&1&0&0\\
                       0&0&1&0\\
                       0&0&0&1 \end{matrix}\right) \mid c\in\mathbf F_q\rbrace.
\end{equation}
The group $Q$ is a normal subgroup of $G_{p_0}$
and $C$ is the center of $Q$, with $Q/C$ elementary 
abelian of order $q^2$.
It is easy to see 
by matrix computations  that $C$ acts trivially on $p_0^\perp$
and that $Q$ stabilizes each line $\ell$ through $p_0$, acting
transitively on the $q$ points of $\ell\setminus\{p_0\}$.
These $q$ points have homogeneous coordinates of the form
$[d:x:y:0]$, where $[x:y]$ are homogeneous coordinates
of a fixed point on a projective line, and $d$ varies over $\mathbf F_q$. It is clear that the subgroup $Q[x:y]$ of index $q$
in  $Q$ consisting of matrices (\ref{Qmatrix}) 
in which $ax+by=0$ is the kernel of the action on 
$\ell\setminus\{p_0\}$ and so $\F[\ell\setminus\{p_0\}]$
affords the regular representation of $Q/Q[x:y]$.

As $[x:y]$ varies over the projective line, we
deduce that, $\F[p_0^\perp\setminus\{p_0\}]$ 
contains the trivial module of $Q/C$ with multiplicity $q+1$. 
Thus since $Q$ has odd order, 
we have a $\F G_{p_0}$-module decomposition
\begin{equation}
\F[p_0^\perp\setminus\{p_0\}]= T\oplus W,
\end{equation}
where $T$ is the $(q+1)$-dimensional space of $Q$-fixed points 
and $W$ has dimension $q^2-1$ and no $Q$-fixed points.
Let $E$ be a splitting field for $Q$ over $\F$, and
consider the action of $G_{p_0}$ on the
characters of $Q/C$ which occur in
$E\otimes_\F W$. Each of the $q^2-1$ nontrivial characters occurs
once. The group of matrices of the form $\diag(\lambda,\mu,\mu^{-1},
\lambda^{-1})$, with $\lambda$, $\mu\in\mathbf F_q\setminus\{0\}$,
lies in $G_{p_0}$ and acts transitively on the $q-1$ nontrivial
elements, hence also on the $q-1$ nontrivial characters,
of each $Q/Q[x:y]$. Then, since $G_{p_0}$
is transitive on the $q+1$ lines through $p_0$, it follows 
that the $q^2-1$ nontrivial characters of $Q/C$ form a single
$G_{p_0}$-orbit. 
%ref for Clifford 
By Clifford's Theorem ((11.1) in \cite{CR:lu})
it follows that $E\otimes_\F W$ is a simple $EG_{p_0}$-module. Hence
$W$ is a simple $\F G_{p_0}$-module.

We are now ready to consider the intersection
\begin{equation}
\ker\pi_{P_1}\cap C(P,L)=\F[p_0^\perp]\cap C(P,L),
\end{equation}
which is an $\F G_{p_0}$-submodule of $\F[p_0^\perp]$.
Clearly, $X$ is a linearly independent subset of this intersection.
Moreover, each element of $X$ is a fixed point of $Q$. 
We must prove that the intersection is no bigger
than the span of $X$. If it were, then by what we know of
the $\F G_{p_0}$-submodules of $\F[p_0^\perp]$, we see
that either $\F[p_0^\perp]\cap C(P,L)$ must contain
all the $Q$-fixed points of  $\F[p_0^\perp]$ or else
it must contain $W$. The first possibilty is ruled out
because it implies that $C(P,L)$ contains the characteristic
function of the point $p_0$, which is absurd since the number
of points on a line is even. In the second case,
we would have that  $\F[p_0^\perp]\cap C(P,L)$ is of codimension
one in  $\F[p_0^\perp]$. Then, for any point $p\in p_0^\perp$, 
since neither $\chi_p$ nor $\chi_{p_0}$ is in $C(P,L)$,
we would  have $\chi_p-\chi_{p_0}\in C(P,L)$.
Then, by transitivity of $\Symp(V)$ on $P$ and the
connectedness of the adjacency graph of $P$, we would have
that $\chi_p-\chi_{p_0}\in C(P,L)$ for all points $p\in P$,
leading to the conclusion that $C(P,L)$ has
codimension one in $\F[P]$, contrary to known fact.
Thus, the intersection is as claimed.
\end{proof}

% needed in second column of first page if using \pubid
%\pubidadjcol

\begin{lemma}\label{q-1} $\ker\pi_{P_1}\cap C(P,L_1)$ has
dimension $q-1$, and basis the set of functions
$\chi_\ell-\chi_{\ell'}$, where $\ell\neq\ell_0$
is an arbitrary but fixed line through $p_0$ and
$\ell'$ varies over the $q-1$ lines through $p_0$
different from $\ell_0$ and $\ell$.
\end{lemma}
\begin{proof} By Corollary~\ref{linediffs} applied
to $p_0$, we see that if   $\ell$ and $\ell'$
are any two of the $q$ lines through
$p_0$ other than $\ell_0$, the function $\chi_\ell-\chi_{\ell'}$
lies in $C(P, L_1)$. It is obviously in $\ker\pi_{P_1}$.
Clearly, we can find $q-1$ linearly independent functions 
of this kind as described in the statement.
Thus $\ker\pi_{P_1}\cap C(P,L_1)$ has
dimension $\geq q-1$. On the other hand $C(P,L_1)$
is in the kernel of the restriction map to $\ell_0$, while
the image of the restriction of $\ker\pi_{P_1}\cap C(P,L)$ in $\ell_0$
has dimension 2, spanned by the images of $\chi_{\ell_0}$ and 
$\chi_{p_0}$. Thus $\ker\pi_{P_1}\cap C(P,L_1)$ has codimension
at least 2 in $\ker\pi_{P_1}\cap C(P,L)$, which has dimension $q+1$,
by Lemma~\ref{kerP}.
\end{proof}

Our final lemma completes the proof of  Theorem~\ref{main}.

\begin{lemma}\label{span} 
$Z\cup X_0\cup Y$ spans $C(P,L)$ as a vector
space over $\F$.
\end{lemma} 
\begin{proof} 
By Lemma~\ref{q-1}, the span of $X_0\cup Z$
is equal to the subspace spanned by $X_0$ and $L_1$, since 
$\ker\pi_{P_1}\cap C(P,L_1)$ is contained in the span of $X_0$.
We must show that the subspace spanned by $X_0\cup Y$ and $L_1$
contains the characteristic functions of all lines intersecting
$\ell_0$,
including $\ell_0$. First, consider a line $\ell\neq\ell_0$ 
meeting $\ell_0$. We can assume that $\ell$ meets $\ell_0$ 
at a point other than $p_0$, since otherwise $\ell\in X_0$.
Therefore $\ell$ meets $\ell_0$ in the same point $p$ as some
element $\ell'\in Y$. Then Corollary~\ref{linediffs} shows
that $\chi_\ell$ lies in the subspace spanned by $Y$ and $L_1$.
The only line still missing is
$\ell_0$, so our last task is to show that 
$\chi_{\ell_0}$ lies in the span of the characteristic
functions of all other lines. 
First, by Lemma~\ref{allone} applied to $\ell_0$,
we see that the constant function 1 is in the span.
Finally, we see from Lemma~\ref{Phi} that 
\begin{equation}
\sum_{\ell\in X_0}\Phi_\ell=1-\chi_{\ell_0},
\end{equation}
so we are done.
\end{proof}

%\section{Concluding remarks}
\begin{remark}
One can also consider the 
binary code $\LU(3,q)$ when $q=2^t$, $t\geq 1$. 
The exact dimension is not known yet, but
Corollary~\ref{LUlowerbound} provides a lower bound,
since by \cite{SastrySin:lu}  we have 
\begin{equation}\label{rad17}
\dim_\F C(P,L) = 1 + {\left(\frac{1+\sqrt{17}}{2}\right)}^{2t}+{\left(\frac{1-\sqrt{17}}{2}\right)}^{2t}.
\end{equation}
%changed here to make point point more clearly.
This formula is quite different from the
one for odd $q$. Nevertheless, it may well be that the inequality
(\ref{ineq}) is an equality for even $q$, just as it is for odd $q$,
despite the difference in the $\dim_\F C(P,L)$ term. 
Computer calculations of J.-L. Kim  verify this up to $q=16$.
\end{remark}
%\appendices
%\section{}
\appendix
In this appendix $q$ is an arbitrary prime power.
Here we  explain  why our incidence system $(P_1,L_1)$
is equivalent to the incidence system $(P^*, L^*)$
defined by the equations (\ref{kimeqn}).
The explanation is given by the classical Klein
correspondence.

We first look at $(P_1,L_1)$ in
coordinates.
Let $x_0$, $x_1$, $x_2$, $x_3$ be  homogeneous coordinates of $P$
corresponding to our symplectic basis.
Recalling that $p_0=\langle e_0\rangle$,
we see that $P_1$ is the set of points such that
$x_3\neq0$. If we represent such a point as
$(a:b:c:1)$ we have a bijection of $P_1$ 
with ${\mathbf F_q}^3$.

Our choice of basis of $V$ yields the  basis
$e_i\wedge e_j$, for $0\leq i<j\leq3$, of the exterior square
$\wedge^2(V)$. 
Denote the corresponding homogeneous
coordinates of the projective space $\mathbf P(\wedge^2(V))$
by $p_{01}$, $p_{02}$, $p_{03}$, $p_{12}$, $p_{13}$ and $p_{23}$.
A $2$-dimensional subspace of $V$ spanned by
vectors $\sum_{i=0}^3 a_ie_i$ and $\sum_{i=0}^3b_ie_i$
defines, by taking its exterior square, a point of
$\mathbf P(\wedge^2(V))$ with coordinates $p_{ij}=a_ib_j-a_jb_i$,
known as the {\it Pl\"ucker} or {\it Grassmann} coordinates of the subspace.
The totality of points of 
$\mathbf P(\wedge^2(V))$ obtained in this way
from lines of $\mathbf P(V)$ forms the set with equation 
$p_{01}p_{23}-p_{02}p_{13}+p_{03}p_{12}=0$,
called the {\it Klein Quadric}.
The totally isotropic $2$-dimensional subspaces of $V$,
namely the lines of $L$, correspond to those points of the
Klein quadric which satisfy the additional linear
equation $p_{03}=-p_{12}$. Recalling that
$\ell_0=\langle e_0, e_1\rangle$, the set $L_1$ is the subset
of $L$ given by $p_{23}\neq 0$, so taking into consideration
the quadratic relation, we see that $L_1$ consists of the points of
$\mathbf P(\wedge^2(V))$ which have Pl\"ucker coordinates
$(z^2+xy:x:z:-z:y:1)$, hence is in bijection with ${\mathbf F_q}^3$.
Next we consider when $(a:b:c:1)\in P_1$ is contained
in $(z^2+xy:x:z:-z:y:1)\in L_1$. Suppose the latter
is spanned by points with homogeneous coordinates
$(a_0:a_1:a_2:a_3)$ and $(b_0:b_1:b_2:b_3)$.
The given point and line are incident if and only if
all $3\times 3$ minors of the matrix
\begin{equation}
\left(\begin{matrix}a & b & c &1\\
                    a_0 & a_1  & a_2 & a_3\\
                    b_0 & b_1 & b_2 & b_3\\
\end{matrix}\right)
\end{equation}
are zero.
The four equations which result reduce to the two equations
\begin{equation}\label{inceqn}
z=-cy+b, \qquad x=cz-a.
\end{equation}

By a simple change of coordinates, these equations
transform to (\ref{kimeqn}). This shows that
$(P_1,L_1)$ and $(P^*,L^*)$ are equivalent.

%\section*{Acknowledgment} 

%}}}
\newpage
%\bibliography{IEEEabrv,final_lu}

\begin{thebibliography}{1}
\providecommand{\url}[1]{#1}
\csname url@rmstyle\endcsname
\providecommand{\newblock}{\relax}
\providecommand{\bibinfo}[2]{#2}
\providecommand\BIBentrySTDinterwordspacing{\spaceskip=0pt\relax}
\providecommand\BIBentryALTinterwordstretchfactor{4}
\providecommand\BIBentryALTinterwordspacing{\spaceskip=\fontdimen2\font plus
\BIBentryALTinterwordstretchfactor\fontdimen3\font minus
  \fontdimen4\font\relax}
\providecommand\BIBforeignlanguage[2]{{%
\expandafter\ifx\csname l@#1\endcsname\relax
\typeout{** WARNING: IEEEtran.bst: No hyphenation pattern has been}%
\typeout{** loaded for the language `#1'. Using the pattern for}%
\typeout{** the default language instead.}%
\else
\language=\csname l@#1\endcsname
\fi
#2}}

\bibitem{Bagchi:lu}
B.Bagchi, A.E.Brouwer, and H.A.Wilbrink, ``Notes on binary codes related to the
  {O}$(5,q)$ generalized quadrangle for odd q,'' \emph{Geometriae Dedicata},
  vol.~39, pp. 339--355, 1991.

\bibitem{Kim:lu}
J.-L. Kim, U.~Peled, I.~Perepelitsa, V.~Pless, and S.~Friedland, ``Explicit
  construction of families of {LDPC} codes with no 4-cycles,'' \emph{{IEEE}
  Trans. Inform. Theory}, vol.~50, pp. 2378--2388, 2004.

\bibitem{LU:lu}
F.~Lazebnik and V.~A. Ustimenko, ``Explicit construction of graphs with
  arbitrarily large girth and of large size,'' \emph{Discrete Applied Math.},
  vol.~60, pp. 275--284, 1997.

\bibitem{CR:lu}
C.~W. Curtis and I.~Reiner, \emph{Methods of Representation Theory, with
  Applications to Finite Groups and Orders}.\hskip 1em plus 0.5em minus
  0.4em\relax New York, NY: Wiley Interscience, 1981, vol.~I.

\bibitem{SastrySin:lu}
N.~S.~N. Sastry and P.~Sin, ``The code of a regular generalized quadrangle of
  even order,'' in \emph{Group Representations: Cohomology, Group Actions and
  Topology}, ser. Proc. Symposia in Pure Mathematics, vol.~63, 1998, pp.
  485--496.

\end{thebibliography}

\begin{biographynophoto}{Peter Sin}
received the B.Sc. degree in mathematics from the 
University of Warwick in 1983 and the D.Phil. degree from the University
of Oxford in 1986. He was an L. E. Dickson Instructor at the University
of Chicago Department of Mathematics from 1986 to 1988. He joined 
the Mathematics Department of the University of Florida 
in 1989 and is now a Professor there. His research interests include
the geometry and representation theory of finite and algebraic groups
and their connections with combinatorics and algebraic coding theory. 
\end{biographynophoto}

% insert where needed to balance the two columns on the last page
%\newpage

\begin{biographynophoto}{Qing Xiang}
 received the Ph.D. degree in mathematics from Ohio State
 University, Columbus, OH, in 1995.

 He was a Harry Bateman research instructor at Caltech, Pasadena, CA, from
 1995 to 1997. Since 1997, he has been on the faculty of Department of
 Mathematical Sciences, University of Delaware, Newark, DE. His research
 interests include combinatorial design theory, algebraic coding theory and
 finite geometry.

 Dr. Xiang was a recipient of the Kirkman Medal from the Institute of
 Combinatorics and its Applications in 1999.
\end{biographynophoto}

\end{document}